\documentclass[aps,prb,reprint,superscriptaddress]{revtex4-1}
\usepackage{mathtools}
\usepackage{amsmath, amssymb, amsthm}
\usepackage{bm}
\usepackage{subfigure}
\usepackage{stmaryrd}
\usepackage{graphicx,tabularx}
\usepackage{hyperref}
\usepackage{bbding}
\begin{document}

\title{Classification and Monte Carlo study of symmetric $\mathbb{Z}_{2}$ spin liquids on the triangular lattice}

\author{Wayne Zheng}
\email[\Envelope~]{intuitionofmind@gmail.com}
\affiliation{Institute for Advanced Study, Tsinghua University, Beijing 100084, China}
\author{Jia-Wei Mei}
\affiliation{Perimeter Institute for Theoretical Physics, Waterloo, ON N2L 2Y5, Canada}
\author{Yang Qi}
\affiliation{Institute for Advanced Study, Tsinghua University, Beijing 100084, China}
\affiliation{Perimeter Institute for Theoretical Physics, Waterloo, ON N2L 2Y5, Canada}

\begin{abstract} We study different ways of symmetry fractionalization in $\mathbb{Z}_{2}$ spin liquids on the triangular lattice. Our classification can be used to identify the symmetry fractionalization in the $\mathbb{Z}_2$ spin liquid reported in recent density-matrix-renormalization-group simulations for $J_1$-$J_2$ spin model on the triangular lattice. We find 64 types of symmetry enriched $\mathbb{Z}_{2}$ spin liquid states on triangular lattice. Besides 8 states constructed in Schwinger-boson parton wavefunctions, 12 more states can be realized in Abrikosov-fermion parton construction. Within a larger gauge group than $SU(2)$, the rest 40 states are also found in a spin-3/2 system. Among 20 types of Abrikosov-fermion Gutzwiller-projected wavefunctions, No.B5 state is a promising candidate for the $\mathbb{Z}_2$ spin liquid for $J_1$-$J_2$ spin model on the triangular lattice. No.B5 lies close to Dirac spin liquid (DSL). However, variational Monte Carlo simulation find that DSL has a good variational energy and $J_1$-$J_2$ spin model cannot open a gap for spinons on top of DSL to stabilize No.B5 state.  \end{abstract}

\maketitle

\section{Introduction}

Together with global symmetries, Long-range entanglement has been playing an important role in characterizing and classifying the many-body wavefunctions. While the former describes various symmetry-breaking phases~\cite{landau}, the long-range entanglement precisely rephrases the intrinsic topological order in the gapped quantum phases of matter\cite{Wen2004,Levin2006,Kitaev2006}, as it always survives under local unitary operations without closing the gap\cite{ChenPRB2010}. This topological order is fully characterized by the quasi-particle (anyon in the bulk) statistics\cite{Laughlin1983,Wilczek1984,Arovas1984,Wen1989,Wen1990,Wen1990a} (and also the chiral central charge if there are chiral gapless edge states\cite{Wen1992,Wen1995}). In two spatial dimensional systems, the anyon statistics and edge chiral central charge (up to $E_8$ bosonic quantum Hall states) is fully described by the modular $S$ and $T$ matrices that generate a projective representation of $SL(2,\mathbb{Z})$ on the degenerate-ground-state Hilbert space on a torus\cite{Keski-Vakkuri1993,Wen1990a,Wen1989,Wen1990,Wen2012,Liu2013,Moradi2014,He2014,Mei2015}.

When both of the global symmetries and long-range entanglements are present, symmetries further enrich the classifications of the gapped many-body wavefunctions with intrinsic topological orders such that anyon excitations in topological ordered states can carry fractionalized quantum numbers of symmetries\cite{Wen2002,Wen2002a,Essin2013}. For example, $\nu=1/3$ Laughlin state has anyons carrying the electric charge $e/3$ which is the fractionalized quantum number of electromagnetic $U(1)$ symmetry.\cite{Laughlin1983} The quantum spin liquid is another concrete example~\cite{AndersonRVB,PhysRevB.40.7133}. In this work we are primarily interested in spin liquids with the $\mathbb Z_2$ topological order. Besides the trivial particle excitation, $\mathbb{Z}_2$ spin liquid supports three types of anyon excitations: the bosonic spinon $b$, the vison $v$ and the fermionic spinon $f$, which is the bound state of $b$ and $v$. Spinons in $\mathbb Z_2$ spin liquid state carry a half-integer spin~\cite{AndersonRVB, PhysRevB.40.7133} representing the fractionalization of SO(3) spin rotation symmetry. Moreover, anyons ($b$, $v$ and $f$) can also carry fractionalized quantum numbers of the crystal symmetry\cite{Wen2002,Wen2002a}, e.g., the anyons can carry half-integer angular momentum or the translations in two different directions can become anticommutative when acting on an anyon\cite{PhysRevB.91.100401, ArXiv14030575,ArXiv150101395}. Different ways of symmetry fractionalization distinguish different types of $\mathbb Z_2$ spin liquid states, and this difference can be detected by experiments and numerical simulations\cite{Wen2002a, WangSET, PhysRevB.91.100401, ArXiv14030575,ArXiv150101395}.

The search for quantum spin liquids is a long sought goal in condensed matter physics\cite{AndersonRVB,Balents2010}. $\mathbb Z_2$ spin liquid states have been discovered by numerical studies in several two-dimensional frustrated spin systems~\cite{Yan2011, PhysRevLett.109.067201, Jiang:2012aa,Hu2013}, and recently in the $J_1$-$J_2$ Heisenberg model on the triangular lattice~\cite{ArXiv150204831, ArXiv150400654}. To explore the symmetry enriched properties, we classify different ways of fractionalizing crystal and time-reversal symmetries on the triangular lattice in $\mathbb Z_2$ spin liquid state. Taking into account two physical constraints that the spinons ($b$ and $f$) carry a half-integer spin and the vison is described by an odd $\mathbb Z_2$ gauge theory~\cite{Sachdev2000, sondhi}, we find that there are in total 64 different ways to fractionalize the crystal and time-reversal symmetries, each of which describes one type of $\mathbb Z_2$ spin liquid state. These 64 states can labeled by six $\mathbb Z_2$ quantum numbers labeling commutation relation fractionalization and quantum number fractionalization~\cite{PhysRevB.91.100401}, and we propose future numerical studies to measure these quantum numbers with methods described in Ref.~\onlinecite{PhysRevB.91.100401, ArXiv14030575,ArXiv150101395}.

The wavefunctions of $\mathbb{Z}_2$ spin liquids can be written in terms of spinon ($b$ or $f$) representations. Actually, a physical spin is fractionalized into either two bosonic spinons ($b$) or two fermionic spinons ($f$), known as the Schwinger boson\cite{Arovas1988} and the Abrikosov fermion\cite{Arovas1988,Baskaran1988}, respectively. We can write down the mean field wavefunctions for bosonic ($b$) and fermionic ($f$) spinons and implement Gutzwiller projection to project out double occupied configurations in the mean field wavefunctions.  The symmetry fractionalization can be studied using the projective symmetry group (PSG) analysis on the Gutzwiller-projected wavefunctions (GPWF) for $\mathbb{Z}_2$ spin liquids \cite{Wen2002,Wen2002a}.  Out of these 64 states, eight has been constructed using Schwinger-boson parton construction in Ref.~\onlinecite{sstri,PhysRevB.74.174423}. We find that using Abrikosov-fermion parton construction with an SU(2) gauge symmetry, 12 more states can be constructed besides these eight states. The connection between Schwinger-boson and Abrikosov-fermion constructions is established for the eight GPWFs. For Abrikosov-fermion GPWFs, variational Monte Carlo method is implemented to search for the candidate $\mathbb{Z}_2$ spin liquid for $J_1$-$J_2$ spin model on the triangle lattice. No.B5 state in TABLE \ref{tab:fermionic_spinon_classification} that lies in the neighbor of U(1) Dirac spin liquid (DSL) is promising. However, our preliminary variational Monte Carlo simulations find that the spinon gap on top of DSL is not favored  and DSL has a good variational energy for the $J_1$-$J_2$ spin model, consistent with previous systematic variational Monte Carlo search\cite{Kaneko2014} where gapless spin liquid is reported. The parton construction only with a SU(2) gauge symmetry doesn't give rise to all possible ways of symmetry fractionalization in $\mathbb{Z}_2$ spin liquids\cite{Wen2002,Wen2002a,Essin2013}. We consider a spin-3/2 system and find out the rest of the 44 states within a gauge symmetry larger than SU(2).

The rest of the paper is organized in the following: In Sec. \ref{sec:02} we use projective method to investigate $\mathbb{Z}_{2}$ spin liquids, including a general consideration ahead, then the classification of 20 kinds of fermionic spinons on triangular lattice and mapping between Schwinger boson states to Abrikosov fermion states bridged by vison's fractionalized quantum numbers derived from a dual $\mathbb{Z}_{2}$ gauge field theory. In the next Sec. \ref{sec:03} we carry out preliminary variational Monte Carlo simulations on some of these states. Enlightened by the discussion in Sec. \ref{sec:02}, in Sec. \ref{sec:04} we propose to construct all the $\mathbb{Z}_{2}$ spin liquid states beyond conventional Abrikosov fermion and Schwinger boson formulations.

\section{Symmetry fractionalizations in $\mathbb{Z}_{2}$ spin liquids}
\label{sec:02}
\subsection{General Consideration}

We begin by classifying different ways to fractionalize crystal and time-reversal symmetries for one type of anyon. In $\mathbb Z_2$ spin liquid, an anyon can carry fractionalized quantum numbers. This is because a physical wave function always contains pairs of anyons (or in general anyons that fuse into a trivial excitation). In presence of symmetries, the wavefunction  carries a linear representation of the symmetry group $G$. Due to fractionalization, anyons carry a projective representation of $G$, which features fractionalized quantum numbers of the global symmetry. Precisely, the relation $f(g)f(h)=f(gh)$ for a linear representation $f$ acting on two group element $g$ and $h$ can be fractionalized into $f(g)f(h)=\omega(g, h)f(gh)$ in a projective representation, where the factor $\omega(g, h)$ labels the symmetry fractionalization.

The symmetry fractionalization embedded in the projective representation must be consistent with the anyon fusion rule. Particularly since two anyons of the same type in $\mathbb{Z}_2$ spin liquid fuse into a trivial excitation, the factor $\omega(g, h)$ can only be $\pm1$. Furthermore, two projective representations that differ by a linear representation represent the same topological order, because we can always attach a trivial local excitation carrying a linear representation to an anyon. Therefore if we consider only one type of anyon, different symmetry enriched topological orders are classified by inequivalent projective representations with $\omega(g, h)=\pm1$, and mathematically this is given by the second group cohomology with $\mathbb Z_2$ coefficient, $\mathcal H^2(G, \mathbb Z_2)$.\cite{Essin2013}

Different elements in $\mathcal H^2(G, \mathbb Z_2)$ are labeled by a set of $\omega$ factors which indicates how an algebraic relation $gh\cdots=1$ in the group algebra of $G$ is promoted to $f(g)f(h)\cdots=\pm1$ in a projective representation. These $\omega$ factors are the cocycle variables in the language of group cohomology. Physically they indicate how the action of symmetry group is fractionalized on the anyon. For example, two commuting symmetry operations $XY=YX$ can become anticommuting $XY=-YX$ when acting on an anyon, and a relation $X^n=1$ can become $X^n=-1$ when acting on an anyon, indicating that the anyon carries a fractionalized quantum number of $X$~\cite{PhysRevB.91.100401}.

\begin{figure}[h!]
\centering
\includegraphics[scale=0.8]{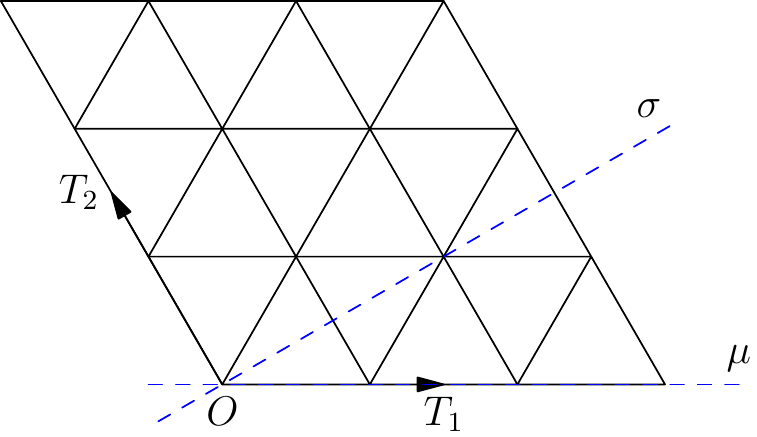}
\caption{Triangular lattice in a oblique coordinate system.}
\label{fig:triangular_lattice}
\end{figure}

On the triangular lattice (Fig. \ref{fig:triangular_lattice}), the symmetry group we consider here is $G=G_{p6m}\times\mathbb Z_2^T$, where $G_{p6m}$ and $\mathbb Z_2^T$ denotes the space group of the lattice and time-reversal, respectively. For this symmetry group, the second group cohomology can be calculated using the HAP package~\cite{HAP1.10.15} in the GAP system~\cite{GAP4}. It can be shown that when calculating the group cohomology, the subgroup of translations can be separated from the point group $C_{6v}$ and the time-reversal symmetry,
\begin{equation}
  \label{eq:p6mc6v}
  \mathcal H^2(G, \mathbb Z_2)
  =\mathbb Z_2\times
  \mathcal H^2(C_{6v}\times\mathbb Z_2^T, \mathbb Z_2),
\end{equation}
where the first $\mathbb Z_2$ denotes the fractionalization of the commutation relation between the two translation operations, $T_1T_2=\pm T_2T_1$. Hence when classifying projective representations on the triangular lattice, the translation symmetry can be considered separately from the point group and time reversal symmetries~\footnote{This separation was shown for the special case of U(1) and SU(2) projective representations in Ref.~\onlinecite{PhysRevB.74.174423, PhysRevB.83.224413}, and their arguments can be easily generalized beyond these special cases.}. (This separation is a special property of the triangular lattice space group and may not apply to other lattices, e.g., the square lattice).

Furthermore, the symmetry fractionalization of the point group and time-reversal symmetries is classified by six more $\mathbb Z_2$ labels as indicated by $\mathcal H^2(C_{6v}\times\mathbb Z_2^T, \mathbb Z_2)=Z_2^6$. In this paper we choose the following six $\mathbb Z_2$ variables $\omega_\sigma$, $\omega_\mu$, $\omega_I$, $\omega_T$, $\omega_{\sigma t}$, $\omega_{\mu t}$, where each $\omega_X$ ($X=\sigma,\mu,I$) labels a quantum number fractionalization~\cite{PhysRevB.91.100401} $X^2=\pm1$ for the symmetry operation $X$, as listed in Table~\ref{tab:algebraic_constraints}. Here $\sigma$ and $\mu$ are two mirror reflection symmetry operations as labeled in Fig.~\ref{fig:triangular_lattice}, $T$ is the time-reversal symmetry and $I$ is the inversion (or 180-degree rotation) symmetry $I=(\sigma\mu)^3$. Together with the translations $T_1$ and $T_2$ they generate the space group $G_{p6m}$ and their action is the following in the coordinate system shown in Fig.~\ref{fig:triangular_lattice},
\begin{equation}
 \begin{split}
 T_{1}: (r_{1}, r_{2})&\rightarrow(r_{1}+1, r_{2}), \\
 T_{2}:(r_{1}, r_{2})&\rightarrow(r_{1}, r_{2}+1), \\
 \mu: (r_{1}, r_{2})&\rightarrow(r_{1}-r_{2}, -r_{2}), \\
 \sigma: (r_{1}, r_{2})&\rightarrow(r_{1}, r_{1}-r_{2}), \\
 I: (r_{1}, r_{2})&\rightarrow(-r_{1}, -r_{2}).
 \end{split}
 \label{eq:triangular_generators}
\end{equation}

The symmetry fractionalization of one type of anyon is classified by the aforementioned seven $\mathbb Z_2$ labels, as $\mathcal H^2(G,\mathbb Z_2)=Z_2^7$. In general each type of anyon in the $\mathbb Z_2$ topological order can have different symmetry fractionalization, but their symmetry fractionalization must be consistent with the anyon fusion rule, so the symmetry fractionalization of two anyons determines the symmetry fractionalization of the third. Moreover, some combinations have obstructions and can only be realized on the surface of a three-dimensional nontrivial symmetry-protected topologically ordered state. Here we further restrict ourselves to $\mathbb Z_2$ spin liquid states that satisfies the following two physical conditions. First, we assume that the vison is described by an odd $\mathbb Z_2$ gauge theory on the dual lattice and therefore has a fixed way of symmetry fractionalization, which is studied in Sec.~\ref{sec:vison}. As a result, there is a one-to-one correspondence between the symmetry fractionalization of bosonic and fermionic spinon. Second, we assume that the spinons carry a half-integer spin and therefore $T^2=-1$. This fixes the label $\omega_T=-1$ for the bosonic and fermionic spinons. With these two constraints, the symmetry fractionalization in a $\mathbb Z_2$ spin liquid state is labeled by six $\mathbb Z_2$ labels, and therefore there are $2^6=64$ different types of spin liquids.

The symmetry fractionalization quantum numbers we list in Table~\ref{tab:algebraic_constraints} can be directly measured using the symmetry quantum numbers and symmetry representation of Schmidt eigenstates in the entanglement spectrum of the ground state wave functions obtained in numerical simulations including the DMRG studies\cite{PhysRevB.91.100401,ArXiv150101395}. First, the commutation relation fractionalization number $\omega_{12}$ can be measured from the ground state momentum in the direction of $T_2$ on a torus with odd number of unit cells in the direction of $T_1$. Second, the fractional quantum number of the crystal symmetries $\omega_\sigma$, $\omega_\mu$ and $\omega_I$ can be measured from the symmetry quantum numbers of the ground state wave functions. Lastly, the fractional quantum numbers $X^2=\omega_X$ can be measured as a one-dimensional SPT invariant by calculating how the symmetry acts on the Schmidt eigenstates in the entanglement spectrum, as all symmetry operations listed in Table~\ref{tab:algebraic_constraints} can be realized either as an on-site antiunitary operation or an unitary and left-right-exchanging operation if we view the system as a quasi-one-dimensional system.

\begin{table}
\centering
\caption{Quantum numbers labeling different projective representations of the symmetry group $\mathcal{H}^2(G, \mathbb Z_2)$.}
\begin{ruledtabular}
\begin{tabular}{>{\centering}p{10mm}>{\centering}p{40mm}>{\centering}p{35mm}}
No. & Algebraic relation & Quantum number\tabularnewline
\hline
1 & $T_1T_2=\pm T_2T_1$ & $\omega_{12}$ \tabularnewline
2 & $\mu^2=\pm1$ & $\omega_\mu$\tabularnewline
3 & $\sigma^2=\pm1$ & $\omega_\sigma$\tabularnewline
4 & $I^2=\pm1$ & $\omega_I$\tabularnewline
5 & $T^2=\pm1$ & $\omega_T$\tabularnewline
6 & $(\mu T)^2=\pm1$ & $\omega_{\mu t}$\tabularnewline
7 & $(\sigma T)^2=\pm1$ & $\omega_{\sigma t}$\tabularnewline
\end{tabular}
\label{tab:algebraic_constraints}
\end{ruledtabular}
\end{table}

\subsection{Symmetry fractionalization in parton constructions of $\mathbb{Z}_{2}$ Spin Liquids} $\mathbb{Z}_2$ spin liquid can be described by GPWFs in parton construction approaches. In these approaches the spin operator is decomposed into two spinons 
\begin{equation}
  \label{eq:slave_particle}
 \bm S_i=\frac12\sum_{\alpha\beta}a_{i\alpha}^\dagger\bm\sigma_{\alpha\beta}
  a_{i\beta}, \quad a_{i\alpha}=f_{i\alpha},b_{i\alpha}
\end{equation}
where $a_{i\alpha}$ can be a fermionic spinon $f_{i\alpha}$ or a bosonic spinon $b_{i\alpha}$ in the Abrikosov fermion and Schwinger boson constructions, respectively. Spinons have the one-particle-per-site constraint $\sum_{\alpha}a_{i\alpha}^{\dagger}a_{i\alpha}=1$. Both parton constructions have an emergent gauge symmetry: In the Schwinger boson construction the physical spin operator in Eq.~\eqref{eq:slave_particle} is invariant under a U(1) symmetry transformation $b_{i\alpha}\rightarrow e^{\text{i}\theta_i}b_{i\alpha}$, whereas in the Abrikosov fermion construction it is invariant under an SU(2) gauge transformation $\psi_i\rightarrow g\psi_i$, where $\psi=
\begin{pmatrix}
  f_{i\uparrow} & f_\downarrow^\dagger
\end{pmatrix}^T$ and $g$ is an element of the SU(2) group generated by Pauli matrices $\tau^{1}, \tau^{2}, \tau^{3}$.

Since we assume that the spinons carry a half-integer spin, $T^2=-1$ is fixed. Hence the spinon transforms under $T$ as $Ta_{i\alpha}=\epsilon_{\alpha\beta}a_{i\beta}$. For other symmetry group actions, in the PSG language the symmetry action $X$ is promoted to $G_XX$, where $G_X$ is an element of the gauge group (SU(2) for Abrikosov fermions and U(1) for Schwinger bosons). Using the generators of the space group, the generic PSG solution for spinons is given as
\begin{equation}
\begin{split}
  &G_{T_1}^{b,f}(r_{1}, r_{2})=\omega_{12}^{r_{2}},\quad G_{T_2}^{b,f}(r_{1}, r_{2})=1,\\
  &G_\mu^{b,f}(r_{1}, r_{2})=\omega_{12}^{r_{1}+r_{2}(r_{2}+1)/2}g_\mu,\\
  &G_\sigma^{b,f}(r_{1}, r_{2})=\omega_{12}^{r_{1}(r_{1}+1)/2}g_\sigma,
\end{split}
\label{eq:psg_solutions}
\end{equation}
where $g_\mu,g_\sigma$ are independent of the site coordinates and satisfy the algebraic relations
\begin{equation}
\begin{split}
&g_\mu^2 = \omega_\mu,\quad g_\sigma^2=\omega_\sigma,\quad (g_\mu g_\sigma)^6=\omega_I,\\
& -g_\mu g_\mu^\ast=\omega_{\mu T}, \quad 
   -g_\sigma g_\sigma^\ast=\omega_{\sigma T}.
\end{split}
\label{eq:psg_relations}
\end{equation}
For fermionic spinons, $g_\mu,g_\sigma\in \text{SU}(2)$ while $g_\mu,g_\sigma$ are $U(1)$ phases for bosonic spinons. Different $\mathbb Z_2$ spin liquids can be constructed by solving these equations with different combinations of symmetry fractionalization quantum numbers. For Abrikosov fermion constructions with an SU(2) gauge symmetry, we find that 20 different spin liquid states can be realized, as listed in Table~\ref{tab:fermionic_spinon_classification}. We note that on the kagom\'e lattice which has the same space group, Abrikosov fermion construction also realizes 20 different states~\cite{PhysRevB.83.224413}, and more states can be realized on an anisotropic triangular lattice~\cite{Zhou2002}, which has a smaller space group. On the other hand \citet{PhysRevB.74.174423} showed that the Schwinger boson construction with U(1) gauge symmetry can realize eight $\mathbb Z_2$ spin liquids states.

We note that neither Schwinger boson nor Abrikosov fermion constructions can realize all 64 different combinations of symmetry fractionalization. This is a restriction due to the gauge symmetry presented in these parton constructions. As an example, in the Abrikosov fermion construction if $\omega_{\mu}=-1$ and $\omega_{\sigma}=+1$, $\omega_{I}$ \emph{must} equal to $-1$ because $g_{\sigma}=\tau^{0}$ without any other choice for the algebra spanned by $\{\tau^{0}, \text{i}\tau^{1}, \text{i}\tau^{2}, \text{i}\tau^{3}\}$ in regard to the two dimensional representation of SU(2) group. The Schwinger boson construction is more restrictive because the U(1) gauge group is smaller than SU(2). The other 44 states that are not constructed in this section requires a parton construction with a gauge symmetry larger than SU(2) and we will give an example of their construction in Sec.~\ref{sec:04}.

The PSG analysis discussed in this section naturally describes the symmetry fractionalization of bosonic spinons and fermionic spinons in the Schwinger boson and Abrikosov fermion constructions, respectively. However, the symmetry fractionalization of the other two types of anyons are not revealed explicitly in this analysis. In the next section we shall see that in all these parton constructions the vison has the same symmetry fractionalization quantum numbers and the symmetry fractionalization of the other type of spinon can be derived by combining the quantum numbers of spinon and vison. Therefore there is a one-to-one correspondence between bosonic and fermionic PSG solutions, and we see that the 20 different spin liquid states we construct using Abrikosov fermion covers the eight states constructed with Schwinger boson. The connection between fermionic and bosonic states is listed in Table~\ref{tab:map_boson_fermion}.


\begin{table}
\centering
\caption{All twenty kinds of Abrikosov fermions on triangular lattice. For states No. A1 - No. A10, $\omega_{12}=1$; for No. B1 - No. B10, $\omega_{12}=-1$.}
\begin{ruledtabular}
\begin{tabular}{>{\centering}p{5mm}>{\centering}p{5mm}>{\centering}p{5mm}>{\centering}p{5mm}>{\centering}p{5mm}>{\centering}p{5mm}>{\centering}p{5mm}>{\centering}p{7mm}>{\centering}p{7mm}>{\centering}p{7mm}>{\centering}p{7mm}>{\centering}p{7mm}}
No. & $\omega_\mu$  & $\omega_\sigma$ & $\omega_{I}$ & $\omega_{\mu t}$ & $\omega_{\sigma t}$  & $g_\mu$ & $g_\sigma$ &$u_\alpha$ & $u_\beta$ & $u_\gamma$& $a_0^l$\tabularnewline
\hline
A1 & $+$ & $+$ & $+$ & $-$ & $-$ & $\tau^0$ & $\tau^0$ & $\tau^{1,3}$ & $\tau^{1,3}$ & $\tau^{1,3}$ & $\tau^{1,3}$\tabularnewline
A2 & $+$ & $-$ & $-$ & $-$ & $+$ & $\tau^0$ & $\text{i}\tau^2$ & $0$ & $0$ & $0$ & $0$\tabularnewline
A3 & $+$ & $-$ & $-$ & $-$ & $-$ & $\tau^0$ & $\text{i}\tau^1$ & $\tau^1$ & $\tau^1$ & $\tau^1$ & $\tau^1$\tabularnewline
A4 & $-$ & $+$ & $-$ & $+$ & $-$ & $\text{i}\tau^2$ & $\tau^0$ & $0$ & $0$ & $0$ & $0$\tabularnewline
A5 & $-$ & $+$ & $-$ & $-$ & $-$ & $\text{i}\tau^1$ & $\tau^0$ & $\tau^1$ & $\tau^1$ & $\tau^1$ & $\tau^1$\tabularnewline
A6 & $-$ & $-$ & $+$ & $+$ & $+$ & $\text{i}\tau^2$ & $\text{i}\tau^2$ & $0$ & $0$ & $0$ & $0$\tabularnewline
A7 & $-$ & $-$ & $+$ & $-$ & $-$ & $\text{i}\tau^1$ & $\text{i}\tau^1$ & $\tau^1$ & $\tau^1$ & $\tau^1$ & $\tau^1$\tabularnewline
A8 & $-$ & $-$ & $-$ & $+$ & $-$ & $\text{i}\tau^2$ & $\text{i}\tau^1$ & $0$ & $0$ & $0$ & $0$\tabularnewline
A9 & $-$ & $-$ & $-$ & $-$ & $+$ & $\text{i}\tau^1$ & $\text{i}\tau^2$ & $0$ & $0$ & $0$ & $0$\tabularnewline
A10 & $-$ & $-$ & $-$ & $-$ & $-$ & $\text{i}\tau^1$ & $\text{i}\tau^3$ & $0$ & $0$ & $0$ & $0$\tabularnewline
B1 & $+$ & $+$ & $+$ & $-$ & $-$ & $\tau^0$ & $\tau^0$ & $0$ & $0$ & $\tau^{1,3}$ & $\tau^{1,3}$\tabularnewline
B2 & $+$ & $-$ & $-$ & $-$ & $+$ & $\tau^0$ & $\text{i}\tau^2$ & $0$ & $\tau^{1,3}$ & $0$ & $0$\tabularnewline
B3 & $+$ & $-$ & $-$ & $-$ & $-$ & $\tau^0$ & $\text{i}\tau^1$ & $0$ & $\tau^3$ & $\tau^1$ & $\tau^1$\tabularnewline
B4 & $-$ & $+$ & $-$ & $+$ & $-$ & $\text{i}\tau^2$ & $\tau^0$ & $\tau^{1,3}$ & $0$ & $0$ & $0$\tabularnewline
B5 & $-$ & $+$ & $-$ & $-$ & $-$ & $\text{i}\tau^1$ & $\tau^0$ & $\tau^3$ & $0$ & $\tau^1$ & $\tau^1$\tabularnewline
B6 & $-$ & $-$ & $+$ & $+$ & $+$ & $\text{i}\tau^2$ & $\text{i}\tau^2$ & $0$ & $0$ & $0$ & $0$\tabularnewline
B7 & $-$ & $-$ & $+$ & $-$ & $-$ & $\text{i}\tau^1$ & $\text{i}\tau^1$ & $0$ & $0$ & $\tau^1$ & $\tau^1$\tabularnewline
B8 & $-$ & $-$ & $-$ & $+$ & $-$ & $\text{i}\tau^2$ & $\text{i}\tau^1$ & $\tau^1$ & $0$ & $0$ & $0$\tabularnewline
B9 & $-$ & $-$ & $-$ & $-$ & $+$ & $\text{i}\tau^1$ & $\text{i}\tau^2$ & $0$ & $\tau^1$ & $0$ & $0$\tabularnewline
B10 & $-$ & $-$ & $-$ & $-$ & $-$ & $\text{i}\tau^1$ & $\text{i}\tau^3$ & $\tau^3$ & $\tau^1$ & $0$ & $0$\tabularnewline
\end{tabular}
\label{tab:fermionic_spinon_classification}
\end{ruledtabular}
\end{table}


\subsection{Vison as a bridge between fermionic and bosonic spinons}
\label{sec:vison}
We start in this subsection to investigate the fractionalization of vison on triangular lattice using a gauge theory and PSG analysis. As a matter of fact, vison is a little bit different from fermionic spinon and bosonic spinon. There is only one kind of vison here thus it is meaningless to talk about  classification. $\omega_{12}^{v}\equiv{-1}$ because a vison sees a spin-1/2 quanta. As mentioned in the introduction section, vison is nothing but the vortex excitation in an \emph{odd} $\mathbb{Z}_{2}$ gauge theory~\cite{Sachdev2000, sondhi}. To begin with a common $\mathbb{Z}_{2}$ gauge theory,
\begin{equation}
H=-J\sum_{}\tau_{l}^{x}-\Gamma\sum_{i}\prod_{l(i)}\tau_{l(i)}^{z},
\label{ising_gauge}
\end{equation}
where variables $\tau^{x}=\pm{1}$ are placed on bonds of triangular lattice and $i$ runs on the dual honeycomb lattice. The conserved Ising gauge charge $\prod_{l(a)}\tau_{l(a)}^{x}=-1$ in an odd Ising gauge theory means the product measurement of all six lines emanating from site $a$, which also can be understood in the quantum dimer model where there is only one dimer on each site\cite{PhysRevB.77.134421}. By defining new operators on the dual honeycomb lattice $\sigma_{i}^{x}=\prod_{l(i)}\tau_{l(i)}^{z}$ and $\sigma_{i}^{z}=\prod_{l<i}\tau_{l}^{x}$ a \emph{Fully Frustrated} Ising Model (FFIM) on dual lattice can be written as\cite{PhysRevB.77.134421,PhysRevB.84.094419,PhysRevB.65.024504} as
\begin{equation}
H_{\text{dual}}=-J_{d}\sum_{\langle{i, j}\rangle}\epsilon_{ij}\sigma_{i}^{z}\sigma_{j}^{z}-\Gamma_{d}\sum_{i}\sigma_{i}^{x},
\label{frustrated_ising}
\end{equation}
in which there is exactly one $\epsilon_{ij}=-1$ along each hexagon, thereby the fractionalized quantum numbers for vison may be calculated from this FFIM. In the dual Ising language, the Ising spins $\sigma^{z}$ are vison creation operators.

\begin{figure}[h!]
\centering
\subfigure[]{
\includegraphics[scale=0.8]{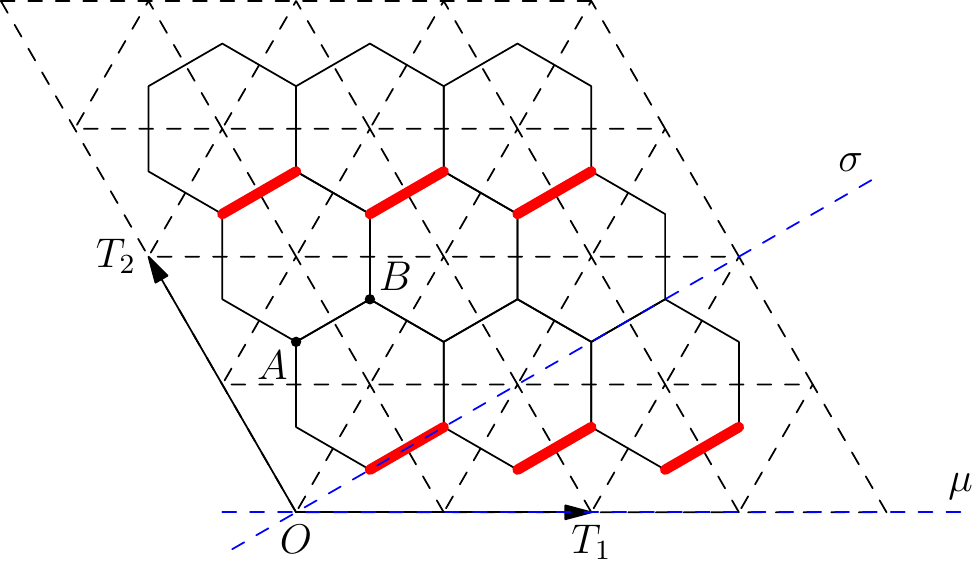}}
\caption{Fully frustrated Ising model on dual honeycomb lattice of triangular lattice. Red and bold bonds represent frustrated ones, namely negative while otherwise are all positive. $A$ and $B$ denote two sublattice for the dual honeycomb lattice.}
\label{fig:dual_ising}
\end{figure}

As shown in Fig. \ref{fig:dual_ising}, the honeycomb lattice has the same space group symmetry as the triangular lattice, which acts in the following way
\begin{equation}
\begin{split}
&T_{1}: (r_{1}, r_{2}, \alpha)\rightarrow(r_{1}+1, r_{2}, \alpha), \\
&T_{2}: (r_{1}, r_{2}, \alpha)\rightarrow(r_{1}, r_{2}+1, \alpha), \\
&\mu: (r_{1}, r_{2}, \alpha)\rightarrow(r_{1}-r_{2}-\frac{1+3\alpha}{6},-r_{2}-\frac{1}{3}, -\alpha), \\
&\sigma: (r_{1}, r_{2}, \alpha)\rightarrow(r_{1}, r_{1}-r_{2}, \alpha), \\
&I: (r_{1}, r_{2}, \alpha)\rightarrow(-r_{1}-\frac{2}{3},-r_{2}-\frac{1}{3}, -\alpha).
\label{eq:vison_transformations}
\end{split}
\end{equation}
where $\alpha=\pm1$ denotes the two sublattices A and B in a unit cell (see Fig.~\ref{fig:dual_ising}). Note that certain symmetry operations exchange the sublattice index in Eq. \ref{eq:vison_transformations}. To be simple and convenient, we define $\tilde{r}_{1}=r_{1}+\frac{1}{3}$ and $\tilde{r}_{2}=r_{2}-\frac{1}{3}$ to represent integer coordinates for hexagons in a same coordinate with triangular lattice. Here we derive a specific PSG solution for vison to compute all the corresponding gauge invariant fractionalized quantum numbers. It is learned that a U(1) gauge symmetry as used in Ref. \onlinecite{PhysRevB.74.174423} is enough to express the state as illustrated in Fig. \ref{fig:dual_ising}. After a bit calculation, we get the explicit PSG solution to vison
\begin{subequations}
\begin{eqnarray}
G_{T_{1}}^{v}(r_{1}, r_{2}, \alpha)&=&1,\quad G_{T_{2}}^{v}(r_{1}, r_{2}, \alpha)=(-1)^{\tilde{r}_{1}}\text{i}^{(1-\alpha)},
\label{eq:vison_psg1} \\
G_{\mu}^{v}(r_{1}, r_{2}, \alpha) 
&=&(-)^{\tilde{r}_{1}+\frac{1}{2}\tilde{r}_{2}(\tilde{r}_{2}+3)}\text{i}^{(1-\alpha)(1+\tilde{r}_2)},
\label{eq:vison_psg3} \\
G_{\sigma}^{v}(r_{1}, r_{2}, \alpha) 
&=&(-)^{\tilde{r}_{1}+\frac{1}{2}\tilde{r}_{1}(\tilde{r}_{1}+1)}\text{i}^{(1+\alpha)\tilde{r}_{1}}, \label{eq:vison_psg4}\\
G_{I}^{v}(r_{1}, r_{2}, \alpha)&=&(-)^{\tilde{r}_{1}}\text{i}^{1+\alpha}.
\label{eq:vison_psg5}
\end{eqnarray}
\end{subequations}
These PSG solutions can then be used to compute the symmetry fractionalization quantum numbers for the vison, and the result is summarized in Table~\ref{tab:map_boson_fermion}.

These results of vison's symmetry fractionalization complements the classification of spinon symmetry fractionalization discussed in the previous section using parton constructions, which only give the symmetry fractionalization of one type of anyon (the Schwinger boson and Abrikosov fermion construction gives the symmetry fractionalization of bosonic spinon and fermionic spinon, respectively). Combining these two results, the symmetry fractionalization of the third type of anyon can be determined. For any quantum number $\omega_X$, the values of the three anyons are related as $\omega_X^f=s_X\omega_X^b\omega_X^v$ because $f$ is a bound state of $b$ and $v$, and $s_X=\pm1$ is a twist factor that arises for symmetry operations that exchanges anyons as a result of the mutual statistics between $b$ and $v$\cite{Essin2013,ArXiv14030575,PhysRevB.91.100401}. Furthermore, the symmetry fractionalization of vison also provides a mapping between symmetry fractionalization quantum numbers of two types of spinons, as summarized in Table~\ref{tab:map_boson_fermion}. The already well known eight kinds of $\mathbb{Z}_{2}$ spin liquids within bosonic construction\cite{PhysRevB.74.174423} can be mapped exactly to part of the states we have found as shown in Table \ref{tab:fermionic_spinon_classification}, where the fractionalized quantum numbers for bosonic states can be obtained explicitly by means of the PSG solutions given in Ref. \onlinecite{PhysRevB.74.174423}. Here we summarize these results in Table \ref{tab:map_boson_fermion_states}. Also note that this mapping can be independently checked using the fact that a nearest-neighbor only bosonic RVB wave function $P_{G}\exp{\left[\xi_{ij}\epsilon_{\alpha\beta}b_{\alpha}^{\dagger}b_{\beta}^{\dagger}\right]}|0\rangle$ can be exactly mapped to a Gutzwiller projected BCS wave function $P_{G}\exp{\left[\zeta_{ij}\epsilon_{\alpha\beta}f_{\alpha}^{\dagger}f_{\beta}^{\dagger}\right]}|0\rangle$\cite{FrusMag,PhysRevLett.109.147209} if $|\xi_{ij}|=|\zeta_{ij}|$ and $\phi_{p}=\phi_{p}^{f}+\pi$, which means the fermionic plaquette flux $\phi_{p}^{f}$ differs $\pi$ from bosonic plaquette flux.


\begin{table}
\centering
\caption{Mapping rules between fermionic states and bosonic states through their fractionalized quantum numbers for $\mathbb{Z}_{2}$ spin liquids on triangular lattice.}
\begin{ruledtabular}
\begin{tabular}{>{\centering}p{10mm}>{\centering}p{15mm}>{\centering}p{10mm}>{\centering}p{30mm}}
fermion & vison & boson & mapping rules \tabularnewline
\hline
$\omega_{12}^{f}$ & $\omega_{12}^{v}=-$ & $\omega_{12}^{b}$ & $\omega_{12}^{b}=\omega_{12}^{f}\cdot\omega_{12}^{v}$ \tabularnewline
$\omega_{\mu}^{f}$ & $\omega_{\mu}^{v}=+$ & $\omega_{\mu}^{b}$ & $\omega_{\mu}^{b}=-\omega_{\mu}^{f}\cdot\omega_{\mu}^{v}$ \tabularnewline
$\omega_{\sigma}^{f}$ & $\omega_{\sigma}^{v}=+$ & $\omega_{\sigma}^{b}$ & $\omega_{\sigma}^{b}=-\omega_{\sigma}^{f}\cdot\omega_{\sigma}^{v}$ \tabularnewline
$\omega_{I}^{f}$ & $\omega_{I}^{v}=-$ & $\omega_{I}^{b}$ & $\omega_{I}^{b}=-\omega_{I}^{v}\cdot\omega_{I}^{v}$ \tabularnewline
$\omega_{\mu{t}}^{f}$ & $\omega_{\mu{t}}^{v}=+$ & $\omega_{\mu{t}}^{b}$ & $\omega_{\mu{t}}^{b}=\omega_{\mu{t}}^{f}\cdot\omega_{\mu{t}}^{v}$ \tabularnewline
$\omega_{\sigma{t}}^{f}$ & $\omega_{\sigma{t}}^{v}=+$ & $\omega_{\sigma{t}}^{b}$ & $\omega_{\sigma{t}}^{b}=\omega_{\sigma{t}}^{f}\cdot\omega_{\sigma{t}}^{v}$ \tabularnewline
\end{tabular}
\label{tab:map_boson_fermion}
\end{ruledtabular}
\end{table}

\begin{table}
\centering
\caption{Mapping from bosonic states to fermionic states for $\mathbb{Z}_{2}$ spin liquids on triangular lattice.}
\begin{ruledtabular}
\begin{tabular}{>{\centering}p{5mm}>{\centering}p{37mm}>{\centering}p{23mm}}
No. & bosonic states as labeled in Ref. \onlinecite{PhysRevB.74.174423} with $(p_{1}, p_{2}, p_{3})$ & fermionic states as in Table \ref{tab:fermionic_spinon_classification}  \tabularnewline
\hline
 1 & $(0, 0, 0)$ & B7 \tabularnewline
 2 & $(1, 0, 0)$ & A7 \tabularnewline
 3 & $(0, 1, 0)$ & B1 \tabularnewline
 4 & $(1, 1, 0)$ & A1 \tabularnewline
 5 & $(0, 0, 1)$ & B5 \tabularnewline
 6 & $(1, 0, 1)$ & A5 \tabularnewline
 7 & $(0, 1, 1)$ & B3 \tabularnewline
 8 & $(1, 1, 1)$ & A3 \tabularnewline
\end{tabular}
\label{tab:map_boson_fermion_states}
\end{ruledtabular}
\end{table}

\section{Variational Monte Carlo results}
\label{sec:03}

In this section, we will implement VMC simulations to search for the candidate $\mathbb{Z}_2$ spin liquid for the $J_1$-$J_2$ spin Heisenberg model on the triangular lattice
\begin{equation}
  \label{eq:heisen}
  H=J_1\sum_{\langle ij\rangle}\bm S_i\cdot\bm{S}_j+J_2\sum_{\langle\langle ij\rangle\rangle}\bm{S}_i\cdot\bm{S}_j,
\end{equation}
where ${\langle ij\rangle}$ and $\langle\langle ij\rangle\rangle$ are for the nearest neighbor (nn) and 2nd nn bonds, respectively. Previously the VMC method has been used to study other spin models on the triangular lattice to search for spin liquids\cite{PhysRevB.72.045105, PhysRevLett.111.157203, PhysRevLett.114.167201, KaiLi2015}. Here we choose to study the $J_1$-$J_2$ as motivated by recent DMRG studies\cite{ArXiv150204831, ArXiv150400654}.

Our variational wave function is obtained from the following mean field Hamiltonian for fermionic spinons 
\begin{equation}
  \label{eq:mf}
  H_\text{MF}=-\sum_{ij}\psi_i^\dag U_{ij}\psi_j+\sum_i\mu^a\psi_i^\dag \tau^a\psi_i
\end{equation}
with the double $\psi=(f_{i\uparrow},f_{i\downarrow}^\dag)^T$ and the mean field parameters $U_{ij}=\langle\psi_i\psi_j^\dag\rangle$. 

Let's consider mean field parameters up to 3rd nn bonds ($u_\alpha$, $u_\beta$ and $u_\gamma$ for nn, 2nn and 3nn). The allowed non-zero mean field parameters are listed in TABLE \ref{tab:fermionic_spinon_classification}. Among these 20 different kinds of spin liquids, only No. A1, B1, B3, B5 can be gapped if only mean field parameters up to 3rd nn bonds are considered. Because of a vanishing mean field parameter on nn bonds, No. B1 and B3 are unlikely to be the ground states of $J_1$-$J_2$ model on the triangular lattice.  No. A1 lies closed to U(1) spin liquid with large spinon fermion surface while No. B5 is in the neighbor of DSL. Without any further parameters, DSL is already a good approximation of the ground state of $J_1$ spin model on the triangular lattice\cite{PhysRevB.73.014519,PhysRevB.74.014408}. Gapped $\mathbb{Z}_2$ spin liquid is most likely to lie in the neighbor of DSL and No.B5 is a promising candidate for the $\mathbb{Z}_2$ spin liquid on the triangular lattice. We will study the different perturbations on top of DSL. Furthermore, the B5 state also corresponds to the zero-flux Schwinger-boson mean field state studied in Refs.~\onlinecite{sstri,PhysRevB.74.174423}. Explicitly, we consider the mean field Hamiltonian \begin{eqnarray}
  H_{\text{DSL}}&=&-\sum_{\langle ij\rangle}\text{sign}(ij)\psi_i^\dag \tau^3 \psi_j+\mu\sum_i\psi_i^\dag\tau^1\psi_i\nonumber\\
&&+\lambda\sum_{\langle\langle\langle ij\rangle\rangle\rangle}\text{sign}(ij)\psi_i^\dag\tau^1\psi_j+h\sum_i\mathbf{n}_i\cdot\mathbf{S}_i.
\end{eqnarray}
where the signs for mean field parameters on nn and 3nn bonds are determined by PSG solutions in Eq. (\ref{eq:psg_solutions}). $\mathbf{n}_i=(\cos(\frac{2\pi}{3}(r_{1}+r_{2}),\sin(\frac{2\pi}{3}(r_{1}+r_{2}))))$ is the coplanar AF spin order varying in the $3\times3$ enlarged unit cell. 

The VMC is carried out on a 12$\times$12 lattice. DSL ($\mu=\lambda=h=0$) has the local spin correlations $3\langle\mathbf{S}_i\cdot\mathbf{S}_j\rangle=-0.5285(2)$ on nn bonds, $3\langle\mathbf{S}_i\cdot\mathbf{S}_j\rangle=0.2135(9)$ on 2nn bond and $3\langle\mathbf{S}_i\cdot\mathbf{S}_j\rangle=0.0044(8)$ on 3nn bonds. DSL is taken as a reference state. Local spin correlation variations are studied on different bonds with increasing $h$, $\mu$ and $\lambda$ to see possible instabilities on top of DSL.

\begin{figure}[t]
  \centering
  \includegraphics[width=0.85\columnwidth]{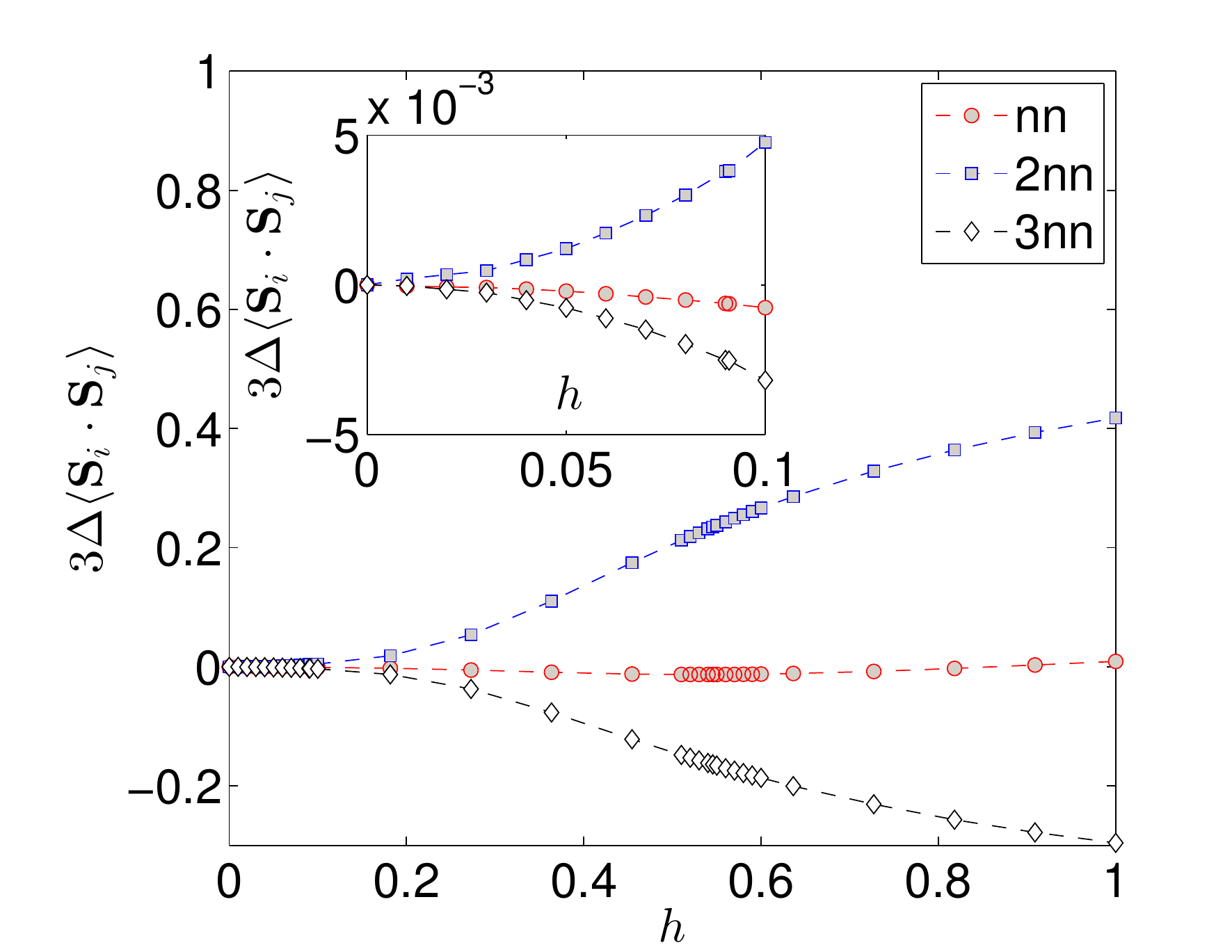}
  \includegraphics[width=0.85\columnwidth]{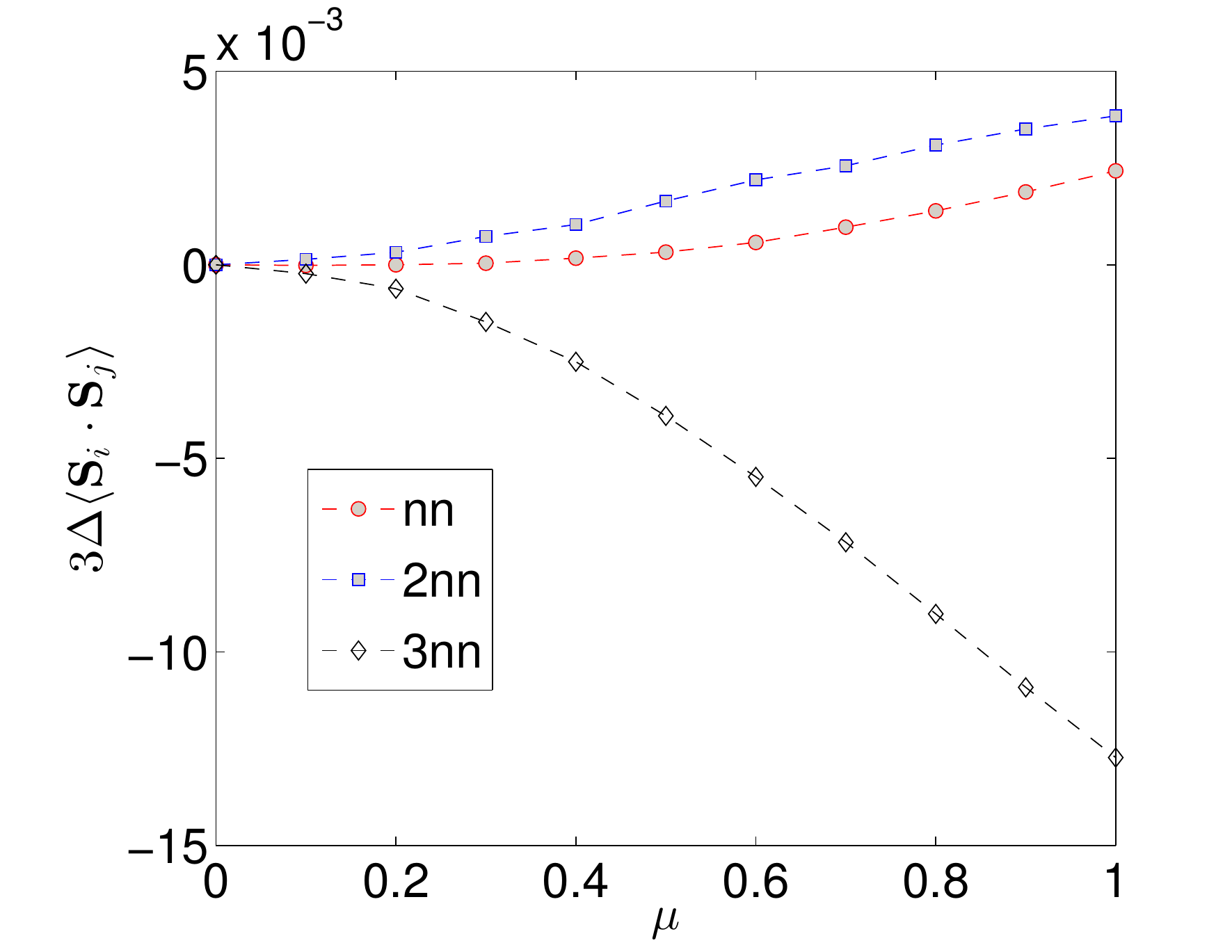}
  \includegraphics[width=0.85\columnwidth]{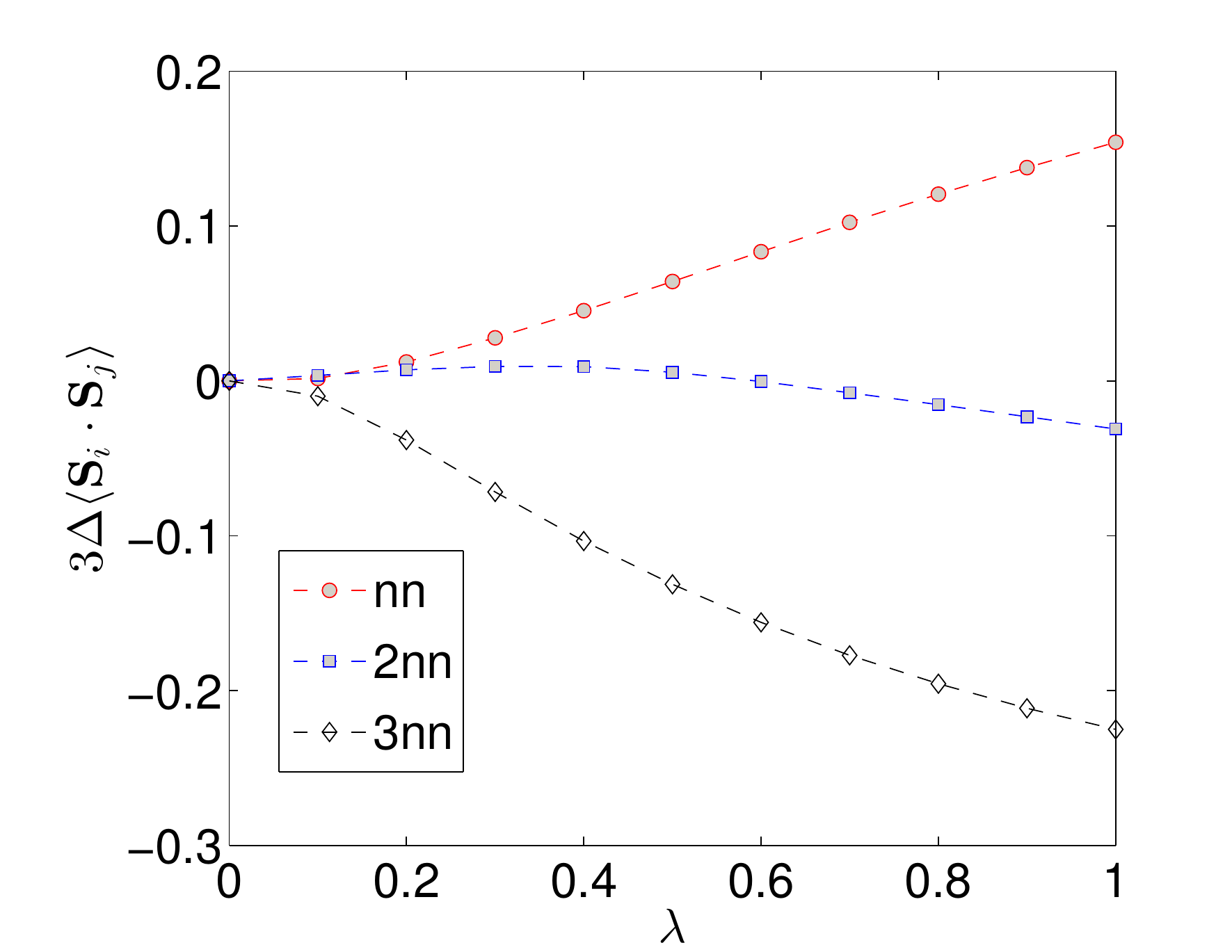}
  \caption{The variational-parameter dependence of local spin correlations on nearest neighbor(nn), 2nd nn (2nn) and 3rd nn (3nn) bonds. DSL is taken as a reference state and $3\Delta\langle\mathbf{S}_i\cdot\mathbf{S}_j\rangle$ is relative local spin correlations. The variational parameters are AF N\'eel spin order $h$ in (a), on-site paring $\mu$ in (b) and 3nn bond hopping amplitude $\lambda$ in (c).}
  \label{fig:DSL}
\end{figure}

In Fig. \ref{fig:DSL} (a), with increasing $h$, the spin correlations on nn bonds decreases implying that  the AF N\'eel spin order is favored by $J_1$ spin model on the triangular lattice. When $h\simeq0.52$, the best variational energy for $J_1$ model is $E_N/(NJ_1)=-0.54142(7)$.\cite{PhysRevB.73.014519,PhysRevB.74.014408} Meanwhile, the increment of spin correlation on 2nd bonds suggests that $J_2$ term will kill the spin order in agreement with DMRG calculations\cite{ArXiv150204831,ArXiv150400654}. 

The on-site chemical potential $\mu$ gaps the Dirac nodes of DSL. With increasing $\mu$, the spin correlations on nn and 2nn bonds increases, but very little as shown in Fig.~\ref{fig:DSL} (b). The bond parameter $\lambda$ on 3nn bonds is also able to open the spinon gap. As shown in Fig. \ref{fig:DSL} (c), when $\lambda$ is small, the spin correlations on nn and 2nd bonds increase with increasing $\lambda$. Therefore, neither $\mu$ or $\lambda$ is favored by $J_1$-$J_2$ spin model on the triangular lattice. The DSL has a good variational energy $E/(NJ_1)=-0.5072(3)$ when $J_2/J_1=0.1$. From these simple attempts, we find that it is not easy to open the spinon gap on top of DSL for $J_1$-$J_2$ model on the triangular lattice. This is consistent with previous VMC simulations in which gapless spin liquid is reported\cite{Kaneko2014}. In addition, we note that similar gapless U(1) spin liquid states are also obtained in VMC studies on the kagom\'e lattice~\cite{PhysRevB.84.020407, PhysRevB.91.020402}, in contrast to the gapped $\mathbb Z_2$ spin liquid state obtained by DMRG studies~\cite{Yan2011, PhysRevLett.109.067201, Jiang:2012aa}.

\section{Realization of $\mathbb{Z}_{2}$ Spin Liquids beyond Schwinger Boson and Abrikosov Fermion Constructions}
\label{sec:04}

As we have demonstrated previously with a concrete example, the Abrikosov fermion construction can not realize all 64 kinds of spin liquid states on triangular lattice due to that the parton construction has only an SU(2) gauge symmetry and the projective representation is restricted in the linear space of an SU(2) fundamental representation. Naturally in this way, this restriction can be lifted and more $\mathbb{Z}_{2}$ spin liquid states are realizable if we consider a larger gauge symmetry in the parton construction, for example, tensor product of two SU(2) fundamental representations. We again use our previously example $\omega_{\mu}=-1, \omega_{\sigma}=+1$ as an illustration. If the gauge group is enlarged to contain products of two SU(2) matrices, $g_{\sigma}$ can be chosen as $g_{\sigma}=(i\tau^{1})\times(i\tau^{1})$ and $g_{\mu}=i\tau^{3}$, therefore resulting in $\omega_{I}=+1$ which is a new state beyond Table~\ref{tab:fermionic_spinon_classification}.

Here we propose a parton construction to realize this kind of exotic
states. Motivated by the work of \citet{ArXiv11104091}, we consider a
spin-$\frac32$ system with three degenerate orbits, each occupied by
one spin-$\frac12$ electron, and we assume that the spins on three
orbits are aligned due to a large Hund's coupling. The spins on the
three orbits can be fractionalized into three sets of Schwinger
fermions,
$\bm S_{ia}=f_{ia\alpha}^\dagger \sigma_{\alpha\beta} f_{ia\beta}$,
where $a=1,2,3$ denotes the orbital degree of freedom.
On the other hand, note also that an even number of half spin within
each unit cell will change the fractionalized quantum numbers of
vison. For illustration, $\omega_{12}^{v}$ will change its sign since
vison travels around a unit cell but does not see net $\mathbb{Z}_{2}$
gauge charge.


Using the Abrikosov fermions, the total spin operator
\begin{equation}
  \label{eq:def_total_spin}
  \bm S_i=\bm S_{i1}+\bm S_{i2}+\bm S_{i3}=
  \sum_{a=1}^3\frac12f_{ia\alpha}^{\dagger}\bm\sigma_{\alpha\beta}f_{ia\beta}
\end{equation}
and the orbital occupation number
\begin{equation}
  \label{eq:total_orb}
  n_a=\sum_{\alpha}f_{ia\alpha}^\dagger f_{ia\alpha}
\end{equation}
are both invariant under the gauge transformation SU(2)$\times$SU(2)$
\times$SU(2), where the three SU(2) subgroups act on the three copies
of Abrikosov fermion operators, respectively. In other words, if we
group the Abrikosov fermion operators as a six-component vector
$\psi_i=\begin{pmatrix}
  f_{i1\uparrow} & f_{i1\downarrow}^\dagger &
  f_{i2\uparrow} & f_{i2\downarrow}^\dagger &
  f_{i3\uparrow} & f_{i3\downarrow}^\dagger
\end{pmatrix}^T$, then the gauge transformation acts as the direct sum
of three SU(2) matrices, $g=g_1\oplus g_2\oplus g_3$.


Next we consider a Abrikosov fermion mean field wave function that breaks the gauge symmetry down to U(1)$\times$U(1)$\times\mathbb Z_2$, where the three subgroups are generated by $(\text{i}\tau^3)\oplus(-\text{i}\tau^3)\oplus1$, ${1}\oplus(\text{i}\tau^3)\oplus(-\text{i}\tau^3)$, and $(-{1})\oplus{1}\oplus{1}$, respectively. Although the invariant gauge group (IGG)~\cite{Wen2002} appears to be larger, this still realizes a $\mathbb Z_2$ spin liquid because a compact U(1) gauge field in (2+1) dimensions is always confined~\cite{Polyakov1977429}. Because of the confinement of the two U(1) groups in the IGG, the remaining gauge group is $\mathcal G=\mathrm{SU}(2)\times \mathrm{SU}(2)\times \mathrm{SU}(2)/ \mathrm U(1)\times\mathrm U(1)$, where we identify $g_1\oplus g_2\oplus g_3$ with $g_1e^{\text{i}\theta\tau^3}\oplus g_2e^{\text{i}(\phi-\theta)\tau^3}\oplus g_3e^{-\text{i}\phi\tau^3}$. Particularly $(-1)\oplus1\oplus1\sim 1\oplus(-1)\oplus1\sim 1\oplus1\oplus(-1)$ is the generator of the $\mathbb Z_2$ IGG. For clarity we denote the element of the gauge group $\mathcal G$ as $\llbracket g_1,g_2,g_3\rrbracket$.


We have constructed all the other 44 kinds of $\mathbb{Z}_{2}$ spin liquid states with the aforementioned gauge group $\mathcal G$ as showed in Table~\ref{tab:exotic_states}. The detailed dynamical properties of these exotic spin liquid states require further investigation in the future.

\begin{table}
\centering
\caption{Exotic $\mathbb{Z}_{2}$ spin liquid states beyond conventional Schwinger boson and Abrikosov fermion constructions on triangular lattice. Listed states No. C1 - No. C22 combined translational fractionalized quantum number $\omega_{12}=\pm{1}$ give birth to all 44 states. Note that in the $\psi$ representation, for example the C1 state, $g_{\mu}$ can be formally written as $g_{\mu}=\oplus_{j=1}^{j=3}g_{\mu}^{j}=\text{i}\tau^{1}\oplus\text{i}\tau^{2}\oplus\tau^{0}$ and correspondingly $\omega_{\mu}=\text{sign}\left[\prod_{j=1}^{3}(g_{\mu}^{j})^{2}\right]=\text{sign}\left(\prod_{j=1}^{3}\omega_{\mu}^{j}\right)$.}
\begin{ruledtabular}
\begin{tabular}{>{\centering}p{7mm}>{\centering}p{5mm}>{\centering}p{5mm}>{\centering}p{5mm}>{\centering}p{5mm}>{\centering}p{5mm}>{\centering}p{20mm}>{\centering}p{20mm}}
 No. & $\omega_{\mu}$ & $\omega_{\sigma}$ & $\omega_{I}$ & $\omega_{\mu{t}}$ & $\omega_{\sigma{t}}$ & $g_{\mu}$ & $g_{\sigma}$ \tabularnewline
\hline
 C1 & $+$ & $+$ & $+$ & $+$ & $+$ & $\llbracket\text{i}\tau^{1}, \text{i}\tau^{2}, \tau^{0}\rrbracket$ & $\llbracket\text{i}\tau^{1}, \text{i}\tau^{2}, \tau^{0}\rrbracket$ \tabularnewline
 C2 & $+$ & $+$ & $+$ & $-$ & $+$ & $\llbracket\text{i}\tau^{1}, \text{i}\tau^{3}, \tau^{0}\rrbracket$ & $\llbracket\text{i}\tau^{3}, \text{i}\tau^{2}, \tau^{0}\rrbracket$ \tabularnewline
 C3 & $+$ & $+$ & $+$ & $+$ & $-$ & $\llbracket\text{i}\tau^{3}, \text{i}\tau^{2}, \tau^{0}\rrbracket$ & $\llbracket\text{i}\tau^{1}, \text{i}\tau^{3}, \tau^{0}\rrbracket$ \tabularnewline
 C4 & $+$ & $-$ & $-$ & $+$ & $+$ & $\llbracket\text{i}\tau^{1}, \text{i}\tau^{2}, \tau^{0}\rrbracket$ & $\llbracket\text{i}\tau^{1}, \text{i}\tau^{2}, \text{i}\tau^{1}\rrbracket$ \tabularnewline
 C5 & $+$ & $-$ & $-$ & $+$ & $-$ & $\llbracket\text{i}\tau^{1}, \text{i}\tau^{2}, \tau^{0}\rrbracket$ & $\llbracket\text{i}\tau^{1}, \text{i}\tau^{2}, \text{i}\tau^{2}\rrbracket$ \tabularnewline
 C6 & $-$ & $+$ & $-$ & $+$ & $+$ & $\llbracket\text{i}\tau^{1}, \text{i}\tau^{2}, \text{i}\tau^{1}\rrbracket$ & $\llbracket\text{i}\tau^{1}, \text{i}\tau^{2}, \tau^{0}\rrbracket$ \tabularnewline
 C7 & $-$ & $+$ & $-$ & $-$ & $+$ & $\llbracket\text{i}\tau^{1}, \text{i}\tau^{2}, \text{i}\tau^{2}\rrbracket$ & $\llbracket\text{i}\tau^{1}, \text{i}\tau^{2}, \tau^{0}\rrbracket$ \tabularnewline
 C8 & $-$ & $-$ & $+$ & $+$ & $-$ & $\llbracket\text{i}\tau^{1}, \text{i}\tau^{2}, \text{i}\tau^{1}\rrbracket$ & $\llbracket\text{i}\tau^{3}, \text{i}\tau^{2}, \text{i}\tau^{2}\rrbracket$ \tabularnewline
 C9 & $-$ & $-$ & $+$ & $-$ & $+$ & $\llbracket\text{i}\tau^{3}, \text{i}\tau^{2}, \text{i}\tau^{2}\rrbracket$ & $\llbracket\text{i}\tau^{1}, \text{i}\tau^{2}, \text{i}\tau^{1}\rrbracket$ \tabularnewline
 C10 & $-$ & $-$ & $-$ & $+$ & $+$ & $\llbracket\text{i}\tau^{1}, \text{i}\tau^{2}, \text{i}\tau^{1}\rrbracket$ & $\llbracket\text{i}\tau^{1}, \text{i}\tau^{2}, \text{i}\tau^{3}\rrbracket$ \tabularnewline
 C11 & $+$ & $+$ & $-$ & $+$ & $+$ & $\llbracket\text{i}\tau^{1}, \text{i}\tau^{2}, \tau^{0}\rrbracket$ & $\llbracket\text{i}\tau^{3}, \text{i}\tau^{2}, \tau^{0}\rrbracket$ \tabularnewline
 C12 & $+$ & $+$ & $-$ & $+$ & $-$ & $\llbracket\text{i}\tau^{1}, \text{i}\tau^{2}, \tau^{0}\rrbracket$ & $\llbracket\text{i}\tau^{1}, \text{i}\tau^{1}, \tau^{0}\rrbracket$ \tabularnewline
 C13 & $+$ & $+$ & $-$ & $-$ & $+$ & $\llbracket\text{i}\tau^{1}, \text{i}\tau^{1}, \tau^{0}\rrbracket$ & $\llbracket\text{i}\tau^{1}, \text{i}\tau^{2}, \tau^{0}\rrbracket$ \tabularnewline
 C14 & $+$ & $+$ & $-$ & $-$ & $-$ & $\llbracket\text{i}\tau^{1}, \text{i}\tau^{1}, \tau^{0}\rrbracket$ & $\llbracket\text{i}\tau^{1}, \text{i}\tau^{3}, \tau^{0}\rrbracket$ \tabularnewline
 C15 & $+$ & $-$ & $+$ & $+$ & $+$ & $\llbracket\text{i}\tau^{3}, \text{i}\tau^{2}, \tau^{0}\rrbracket$ & $\llbracket\text{i}\tau^{1}, \text{i}\tau^{2}, \text{i}\tau^{1}\rrbracket$ \tabularnewline
 C16 & $+$ & $-$ & $+$ & $+$ & $-$ & $\llbracket\text{i}\tau^{3}, \text{i}\tau^{2}, \tau^{0}\rrbracket$ & $\llbracket\text{i}\tau^{1}, \text{i}\tau^{2}, \text{i}\tau^{2}\rrbracket$ \tabularnewline
 C17 & $+$ & $-$ & $+$ & $-$ & $+$ & $\llbracket\text{i}\tau^{1}, \text{i}\tau^{1}, \tau^{0}\rrbracket$ & $\llbracket\text{i}\tau^{1}, \text{i}\tau^{2}, \text{i}\tau^{1}\rrbracket$ \tabularnewline
 C18 & $+$ & $-$ & $+$ & $-$ & $-$ & $\llbracket\text{i}\tau^{1}, \text{i}\tau^{1}, \tau^{0}\rrbracket$ & $\llbracket\text{i}\tau^{1}, \text{i}\tau^{2}, \text{i}\tau^{2}\rrbracket$ \tabularnewline
 C19 & $-$ & $+$ & $+$ & $+$ & $+$ & $\llbracket\text{i}\tau^{1}, \text{i}\tau^{2}, \text{i}\tau^{1}\rrbracket$ & $\llbracket\text{i}\tau^{3}, \text{i}\tau^{2}, \tau^{0}\rrbracket$ \tabularnewline
 C20 & $-$ & $+$ & $+$ & $+$ & $-$ & $\llbracket\text{i}\tau^{1}, \text{i}\tau^{2}, \text{i}\tau^{1}\rrbracket$ & $\llbracket\text{i}\tau^{1}, \text{i}\tau^{1}, \tau^{0}\rrbracket$ \tabularnewline
 C21 & $-$ & $+$ & $+$ & $-$ & $+$ & $\llbracket\text{i}\tau^{1}, \text{i}\tau^{2}, \text{i}\tau^{2}\rrbracket$ & $\llbracket\text{i}\tau^{3}, \text{i}\tau^{2}, \tau^{0}\rrbracket$ \tabularnewline
 C22 & $-$ & $+$ & $+$ & $-$ & $-$ & $\llbracket\text{i}\tau^{1}, \text{i}\tau^{2}, \text{i}\tau^{2}\rrbracket$ & $\llbracket\text{i}\tau^{1}, \text{i}\tau^{1}, \tau^{0}\rrbracket$ \tabularnewline
\end{tabular}
\label{tab:exotic_states}
\end{ruledtabular}
\end{table}

\section{Conclusion}

In this paper, we first studied the symmetry fractionalization in gapped $\mathbb{Z}_{2}$ spin liquids on the triangular lattice, which is described mathematically by the projective representations of translation group, point group and time reversal symmetry groups. There are six independent gauge invariant fractionalized $\mathbb{Z}_{2}$ quantum numbers to label in total 64 kinds of spin liquid states, which can be both illustrated using group cohomology and detailed analysis for triangular lattice's symmetry group. Then we explicitly found that there are 20 spin liquids can be constructed within Abrikosov fermion formulation, while 8 kinds can be constructed by Schwinger boson methods. Simultaneously, we derive vison's fractionalized quantum numbers from another independent way by which vison in $\mathbb{Z}_{2}$ spin liquids can be mapped to a fully frustrated Ising model on the dual lattice. Then we mapped all the bosonic states to fermionic states with the help of anyon fusion rules discussed previously~\cite{ArXiv14030575} to give a unified perspective towards parton constructions of gapped spin liquid states.

Then we have carried out preliminary VMC investigation on these states. We found that for $J_{2}/J_{1}\simeq{0.1}$ the spin order will be killed. However, in contrast to claimed discovery of gapped spin liquids by DMRG recently, we found that the Dirac spin liquid is stable for the $J_{1}-J_{2}$ model on triangular lattice, and the spinon gap does not opened.

In the process of constructing spin liquid states with slave particle approaches we found that the essential constraint comes from the projective point gauge group, of which fermionic construction is SU(2) while bosonic construction is U(1). Motivated by this viewpoint, we propose a method to construct all the other 44 kinds of spin liquid states beyond conventional Schwinger boson and Abrikosov fermion formulations by enlarging the point gauge group to SU(2)$\times$SU(2)$\times$SU(2)/U(1)$\times$U(1). We also discuss one possiable realization in detail by constructing three orbitals enbedded six spin-1/2 fermionic spinons to form the total spin operator within each unit cell. The dynamical properties of these exotic spin liquid states are left to future investigation.

\begin{acknowledgements}
  We thank Zheng-Cheng Gu, Wen-Jun Hu, Zheng-Xin Liu,
  Donna N. Sheng, Qing-Rui Wang and Zhenyue Zhu for invaluable
  discussions. W.Z. is supported by the Scholarship for Overseas Graduate Studies of Tsinghua University to visit Perimeter Institute for Theoretical Physics. Y.Q. is supported by National Basic Research Program of
  China through Grant No. 2011CBA00108 and by NSFC Grant
  No. 11104154. This research was supported in part by
  Perimeter Institute for Theoretical Physics. Research at Perimeter
  Institute is supported by the Government of Canada through Industry
  Canada and by the Province of Ontario through the Ministry of
  Research and Innovation.

  \emph{Note added.} After completing our manucript we were informed
  of a related work~\cite{LuTriPSG}.

\end{acknowledgements}

\bibliographystyle{apsrev4-1}
%
\end{document}